\pdfoutput=1
\documentclass[a4paper,11pt]{article}
\usepackage{graphicx}
\usepackage{fullpage}
\usepackage{framed}

\usepackage{mystyle} 
\graphicspath{{./figures/}}

\usepackage[table]{xcolor}
\usepackage{booktabs}
\usepackage{multirow}

\usepackage{pdflscape}

\usepackage{tikz}
\usetikzlibrary{decorations.pathmorphing, decorations.markings, calc, fadings,
decorations.pathreplacing, patterns,shapes.arrows,positioning,3d,arrows,decorations.markings,shadings,backgrounds}

\usepackage[displaymath, mathlines]{lineno}
\usepackage{framed}
\usepackage{url}
\usepackage{caption,subcaption}

\title{\sffamily Gravitational waves from a  supercooled electroweak phase transition and their detection with pulsar timing arrays}
\author{Archil Kobakhidze$^1$, Cyril Lagger$^1$, Adrian Manning$^1$ and Jason Yue$^{2}$}
\affiliation[]{$^1$ARC Centre of Excellence for Particle Physics at the Terascale, School of Physics, The University of Sydney, NSW 2006, Australia\\
$^2$Department of Physics, National Taiwan Normal University, Taipei 116, Taiwan}

\emailAdd{archil.kobakhidze@sydney.edu.au}
\emailAdd{cyril.lagger@sydney.edu.au}
\emailAdd{adrian.manning@sydney.edu.au}
\emailAdd{jason.yue@ntnu.edu.tw}

\abstract{
We investigate the properties of a stochastic gravitational wave background produced by a first-order electroweak phase transition in the regime of extreme supercooling. We study a scenario whereby the percolation temperature that signifies the completion of the transition, $T_p$, can be as low as a few MeV  (nucleosynthesis temperature), while most of the true vacuum bubbles are formed much earlier at the nucleation temperature, $T_n\sim 50$ GeV. This implies that the gravitational wave spectrum is mainly produced by the collisions of large bubbles and characterised by a large amplitude and a peak frequency as low as $f \sim 10^{-9}-10^{-7}$ Hz. We show that such a scenario can occur in (but not limited to) a model based on a non-linear realisation of the electroweak gauge group, such that the Higgs vacuum configuration is altered by a cubic coupling. In order to carefully quantify the evolution of the phase transition of this model over such a wide temperature range, we go beyond the usual fast transition approximation, taking into account the expansion of the Universe as well as the behaviour of the nucleation probability at low temperatures. Our computation shows that there exists a range of parameters for which the gravitational wave spectrum lies at the edge between the exclusion limits of current pulsar timing array experiments and the detection band of the future Square Kilometre Array observatory.
}

\begin{document} 
\maketitle

%%%%%%%%%%%%%%%%%%%%%%%%%%%%%%%%%%%%%%%%%%%%%
%Section
%%%%%%%%%%%%%%%%%%%%%%%%%%%%%%%%%%%%%%%%%%%%
\section{Introduction}\label{sec:intro}

Cosmological phase transitions (PT) are predicted by many particle physics models with important consequences on the dynamics of the Universe. In particular, first-order PTs produce a stochastic background of gravitational waves (GWs) from the collision of true vacuum bubbles and their interaction with the surrounding hot plasma \cite{Turner:1990rc,Kosowsky:1991ua,Kosowsky:1992rz,Kosowsky:1992vn,Kamionkowski:1993fg}. The observation of a stochastic GW spectrum would then provide an opportunity to obtain more information on the early Universe and potentially new physics. In particular, the Standard Model itself does not predict a first-order electroweak PT \cite{Kajantie:1996mn} and thus neither the production of the associated GWs. However, several extensions of the SM do accommodate such a transition, allowing the matter-antimatter asymmetry of the Universe  to be explained via electroweak baryogenesis  \cite{Kuzmin:1985mm} (cf.~e.g.~\cite{Trodden:1998ym,Morrissey:2012db,Konstandin:2013caa,White:2016nbo} for recent reviews on the topic).

The peak frequency of a stochastic GW background produced by a PT near the electroweak scale, $T_{\text{EW}}\sim 100$ GeV, is expected to lie in the millihertz range, which coincides with the  projected sensitivity of the future eLISA space-based interferometer \cite{Caprini:2015zlo}. This motivated a series of investigations into the production of GWs in various BSM models, see e.g. \cite{Delaunay:2007wb,Kehayias:2009tn,Leitao:2012tx,Leitao:2015fmj,Kakizaki:2015wua,Hashino:2016rvx,Hashino:2016xoj,Huang:2016odd,Vaskonen:2016yiu,Chao:2017vrq,Balazs:2016tbi,Beniwal:2017eik,Cai:2017cbj,Kobakhidze:2016mch}. The characteristic frequency and amplitude of the spectrum are derived from the dynamics of the PT and depend on a few key parameters: the duration of the transition, the size of colliding bubbles, the bubble-walls velocity and the fraction of vacuum energy transferred into the bubble-walls. In the aforementioned studies, these quantities are computed under the assumption that the PT occurs on a time scale much shorter than the Hubble time. The instant at which most of the bubbles are nucleated is thus very close to the time when they collide and cover a significant volume of the Universe. In this article, we shall however consider the case of a prolonged electroweak PT with a non-negligible amount of time between nucleation and collision. We expect the GW background to be predominantly produced by large bubbles colliding much later than in a typical electroweak PT previously discussed in the literature. A range of lower peak frequencies is then observed in such a scenario.

In order to illustrate this phenomenon, we consider a model based on a non-linearly realised electroweak gauge group \cite{Binosi:2012cz, Kobakhidze:2012wb,Kobakhidze:2015xlz,Kobakhidze:2016mch} where the Higgs potential admits a cubic term at tree-level. In other words, a barrier exists between the two different phases of the Higgs field from the electroweak scale down to zero temperature, allowing a significant amount of supercooling. Interestingly, this model exhibits a range of parameters for which the PT is long-lasting, meaning that most of the true vacuum bubbles are nucleated around $T \sim 50$ GeV but collide well below the electroweak scale, as low as $T \sim  [0.1,10]$ GeV. Precise results depend on the exact equation of state of the Universe which is complicated to compute in this context. Naively, vacuum energy is expected to dominate over radiation energy below a given temperature, potentially leading to an inflationary stage. However, we shall argue that such a scenario is unlikely to happen as a significant amount of bubbles are produced early enough (during the radiation dominated era) and subsequently act as a source of inhomogeneity which prevents inflation to occur. Therefore, our model differs from previous studies of scale-invariant models \cite{Randall:2006py,Konstandin:2010cd,Konstandin:2011dr} in which the nucleation of true vacuum bubbles occurs at very low temperatures, namely after inflation started.  We should note that extreme supercooling \cite{Witten:1980ez, Arunasalam1} is also a feature of another class of scale-invariant models of electroweak symmetry breaking with a very light scalar particle \cite{Kobakhidze:2017eml}.

Once the details of the PT are known, rough estimates of the peak frequency and peak amplitude of the GW spectrum can be derived from dimensional arguments, although more precise predictions usually require numerical analysis. Since significant supercooling occurs in our case, bubble collisions become the dominant source of GWs, whilst the interactions of bubbles with the surrounding plasma can be neglected. We then rely on previous numerical simulations employing the so-called envelope approximation \cite{Kosowsky:1991ua,Kosowsky:1992rz,Kosowsky:1992vn,Kamionkowski:1993fg,Caprini:2007xq,Huber:2008hg}. It is important to notice that these simulations have been performed under the assumption of a rapid PT and it is not clear a priori whether they are applicable to longer transitions. However, we take a special care to include the effects of the expansion of the Universe on the related growth of the vacuum bubbles and to study the behaviour of the nucleation probability at low temperature. Together with the previous simulations, this approach is expected to yield good estimates of the GW characteristics. Our calculations show that frequencies of the GWs produced by supercooled phase transitions are in the range $10^{-9}-10^{-7}$ Hz and allows for detection by pulsar timing array (PTA) experiments\footnote{Note also that if the QCD transition were first-order, it could be probed by PTA detectors as well \cite{Caprini:2010xv,Anand:2017kar}. }, such as the future Square Kilometre Array (SKA) \cite{5136190}. 

This article is organised as follows. In Sec.~\ref{sec:model}, we briefly present the model based on a non-linear realisation of the electroweak gauge group. In Sec.~\ref{sec:fin_temp}, we describe the dynamics of a supercooled and long-lasting phase transition. Then we apply this formalism to the aforementioned model. In Sec.~\ref{sec:gw}, we estimate the GW spectrum produced by this phase transition and compare the results with current and future PTA detectors. Finally, we discuss our results and approximations in the conclusion.

%%%%%%%%%%%%%%%%%%%%%%%%%%%%%%%%%%%%%%%%%%%%%%
%%Section
%%%%%%%%%%%%%%%%%%%%%%%%%%%%%%%%%%%%%%%%%%%%%%
\section{A non-linearly realised electroweak gauge group}\label{sec:model}
It has been recently shown that a model with a non-linearly realised electroweak gauge group can accommodate a first-order phase transition \cite{Kobakhidze:2016mch}. We emphasise that the aim of this article is not to focus on the specifics of this model. Rather, such a ``toy-model''  allows us to illustrate a realistic mechanism that is capable of producing GWs associated with a supercooled and long-lasting PT. Henceforth,  a brief description of the key features of this model is given below (cf.~ e.g.~\cite{Binosi:2012cz, Kobakhidze:2012wb,Kobakhidze:2015xlz} for further details).

In this approach, the coset group $\mathcal{G}_\text{coset}=SU(2)_L\times U(1)_Y/U(1)_{Q}$ is gauged and the Higgs boson appears as a singlet $\rho(x) \sim(\mathbf{1},\mathbf{1})_{0}$ under the SM gauge group. The SM model Higgs doublet can then be identified as:     
\begin{equation}
\label{eq:higgs_field_non_linear}
H(x)=\frac{\rho(x)}{\sqrt{2}}e^{\frac{i}{2}\pi^i(x)T^i}\begin{pmatrix}
	0 \\ 
	1 \\ 
\end{pmatrix},\qquad  i\in\{1,2,3\}
\end{equation}
where the three would-be Goldstone bosons spanning the coset space $\mathcal{G}_\text{coset}$ are represented by the $\pi^i (x)$ fields. The broken generators associated with this coset group are correspondingly $T^i=\sigma^{i}-\delta^{i3}\mathbb{I}$, with $\sigma^i$ denoting the Pauli matrices. The physical Higgs $h$ is then identified as the fluctuation of $\rho$ around the electroweak vacuum expectation value $v=246$ GeV such that $\rho = v + h$.

Following \cite{Kobakhidze:2016mch},  all the SM configurations are assumed, with the exception of the Higgs potential. Indeed, as $\rho$ is a singlet, an anomalous cubic term is allowed in the following way:
\begin{equation}\label{eq:potential_tree_level}
V^{(0)}(\rho)=-\frac{\mu^2}{2}\rho^2+\frac{\kappa}{3} \rho^3+\frac{\lambda}{4}\rho^4.
\end{equation} 
This tree-level potential explicitly depends on the three parameters $\mu$, $\kappa$ and $\lambda$. However, the relations $\left.\frac{dV}{d\rho} \right|_{\rho=v}=0$ and $\left.\frac{d^2V}{d\rho^2} \right|_{\rho=v}=m_h^2 \approx (125 \text{ GeV})^2$ allow the model to be controlled by a single free parameter, which is chosen to be $\kappa$. Taking the example at tree level, the above relations can be solved analytically giving:
\begin{equation}\label{eq:vev_relation_tree_level}
  \begin{aligned}
\mu^2& =\frac{1}{2}\left( m_h^2+v\kappa\right),   \\
\lambda &=\frac{1}{2v^2}\left( m_h^2-v\kappa\right).
  \end{aligned}
\end{equation}   
The same process can be used to express $\mu$ and $\lambda$ as a function of $\kappa$ consistently at each order of perturbation, at least numerically. In this article, we solve the relations at one-loop level.

In order to describe the behaviour of the Higgs field in the early Universe, we require the one-loop finite temperature potential. It is usually written as follows \cite{Carrington:1991hz,Liu:1992tn,Dine:1992wr,Arnold:1992rz}:
\begin{equation}\label{eq:finite_potential}
  V(\rho,T)=  V^{(0)}(\rho)+ V^{(1)}_{CW}(\rho) + V^{(1)}(\rho,T) + V_{Daisy}(\rho,T),
\end{equation}
where $V^{(0)}$ is the classical potential (\ref{eq:potential_tree_level}), $V_{CW}^{(1)}$ is the one-loop Coleman-Weinberg potential at $T=0$, $V^{(1)}(\rho,T)$ is the finite temperature contribution and $V_{Daisy}$ are correction terms dealing with infrared divergences. The explicit expressions of these contributions are given in the appendix \ref{app:fin_temp_pot}. For each value of $\kappa$, the potential (\ref{eq:finite_potential}) can be numerically computed and the thermal behaviour of the Higgs field can be analysed.

%%%%%%%%%%%%%%%%%%%%%%%%%%%%%%%%%%%%%%%%%%%%%%
%%Section
%%%%%%%%%%%%%%%%%%%%%%%%%%%%%%%%%%%%%%%%%%%%%%

\section{Prolonged electroweak phase transition}\label{sec:fin_temp}
The cosmological behaviour of the Higgs field is mainly described by its free energy density $\mathcal{F}(\rho, T)=V(\rho, T)$ identified as the effective potential (\ref{eq:finite_potential}). We can summarise its dynamics as a first-order PT using a few key temperatures ($\tilde{T} > T_c > T_n > T_p$) as follows. For $T>\tilde{T}$, $\mathcal{F}(\rho, T)$ admits a single minimum at $\rho=v_{T}^{(+)}$ called the symmetric phase of the Higgs field. As the Universe cools and reaches $\tilde{T}$, a second minimum, called the broken phase, forms at $\rho=v_{T}^{(-)}$  with a free energy density initially higher than that of the symmetric phase. This free energy density then decreases until the two vacua become degenerate at the critical temperature $T=T_c$. For $T<T_c$, the free energy density of the broken phase keeps decreasing, causing the symmetric phase to become metastable: in other words, the Higgs field may tunnel through the potential barrier between $v_{T}^{(+)}$ and $v_{T}^{(-)}$. If the decay probability is high enough, bubbles of true vacuum nucleate and expand in the surrounding symmetric phase. The nucleation temperature, $T_n$ is then defined  as the temperature at which most of the bubbles are produced. On the other hand, the percolation temperature, $T_p$ corresponds to the instant when a significant volume of the Universe (whose value would be specified later on) has been converted from the symmetric to the broken phase. We expect most of bubble-collisions to occur around $T_p$ and not $T_n$, unlike typical high-temperature (short-lived) phase transitions.

We should mention that some models admit a temperature $T_0<T_c$ below which the barrier between the two vacua disappears. If a transition has not occurred by this time, the Higgs field will then roll down the potential without forming any bubbles. This situation does not occur in our model of interest since the potential (\ref{eq:potential_tree_level}) admits a cubic term at zero temperature. In other words, the barrier will never vanish and a first-order PT can occur a priori at arbitrarily low $T$, unless the Higgs field stays trapped in its metastable state. Although an electroweak PT is usually assumed to be quick, with $T_p \sim T_n$, we can in this case consider a longer transition with $T_p \ll T_n$ and a large amount of supercooling. We now show how we can explicitly compute these temperatures from both the nucleation probability and the bubble dynamics.

\subsection{Decay probability}\label{sub:decay_prob}

The tunnelling of the Higgs field between the two vacua is characterised by the decay probability $\Gamma$ per unit time per unit volume. Quantum fluctuations drive this process at zero temperature \cite{Coleman:1977py,Callan:1977pt} while thermal fluctuations dominate at finite $T$ \cite{Linde:1981zj}. Therefore, $\Gamma$ is expressed as a function of the temperature of the Universe and can be written in the semiclassical approximation as follows:
\begin{equation}\label{eq:decay_prob_general}
  \Gamma(T) \approx A(T)  e^{-S(T)}
\end{equation}
where $A(T)$ is a prefactor of mass dimension $4$ and $S(T)$ is the Euclidean action $S[\rho,T]$ evaluated along the bounce trajectory $\rho_B$. In full generality, the Euclidean action is the functional over the Higgs field $\rho$ defined as \cite{Salvio:2016mvj}:
\begin{equation}\label{eq:S_action}
  S[\rho,T]=4\pi \int_{0}^{\beta} d\tau \int^\infty_0 dr\ r^2 \left[  \frac{1}{2}\left( \frac{d\rho}{d\tau} \right)^2+\frac{1}{2}\left( \frac{d\rho}{dr} \right)^2+ \tilde{\mathcal{F}}(\rho,T) \right], 
\end{equation}
where $\tau= -it$ is the Euclidean time, $\beta=\frac1{T}$ and $\tilde{\mathcal{F}}(\rho,T):=V(\rho,T)-V\Big(v_{T}^{(+)},T\Big)$ is the free energy density normalised according to its value in the unbroken phase. The bounce trajectory $\rho_B(\tau, r)$ is the solution which minimises the Euclidean action and thus satisfies the following equation of motion:
\begin{equation}  \label{eq:bounce_general}
  \frac{\partial^2 \rho}{\partial \tau^2}+\frac{\partial^2\rho}{\partial r^2}+\frac{2}{r}\frac{\partial \rho }{\partial r}-\frac{\partial \tilde{\mathcal{F}}}{\partial \rho}(\rho,T)=0,
\end{equation}
with the boundary conditions
\begin{equation}
  \label{eq:bounce_boundary_general}
  \left.\frac{\partial\rho}{\partial\tau}\right|_{\tau=0,\pm \beta/2}=0, \qquad \left.\frac{\partial\rho}{\partial r}\right|_{r=0}=0, \qquad \lim_{r\rightarrow \infty}  \rho(r)= v_{T}^{(+)}.
\end{equation}

The specific shape of the bounce $\rho(\tau, r)$ depends on the temperature \cite{Linde:1981zj}. At zero or low temperature, it reduces to an $O(4)$ symmetric solution $\rho(\tilde{r})$ with  $\tilde{r}=\sqrt{\tau^2+r^2}$, while at high temperature it is given by an $O(3)$-symmetric and time-independent solution $\rho(r)$. The temperature scale that allows us to distinguish between these regimes is given by the mass scale of the problem or equivalently by the size $R_0$ of the $O(4)$ symmetric bubble at $T=0$. In both the limits $T \ll R_0^{-1}$ and $T \gg R_0^{-1}$, the action (\ref{eq:S_action}) simplifies as follows:
\begin{equation}\label{eq:all_actions}
\arraycolsep=1.4pt\def\arraystretch{2.0}
 S[\rho,T] \approx \left\{\begin{array}{llrl}
  S_4[\rho, T] &= &\displaystyle 2\pi^2 \int^\infty_0 d\tilde{r}\ \tilde{r}^3 \left[ \frac{1}{2}\left( \frac{d\rho}{d\tilde{r}} \right)^2+ \tilde{\mathcal{F}}(\rho,T) \right],& T \ll R_0^{-1} \\
   \frac{1}{T}S_3[\rho,T] & = & \displaystyle \frac{4\pi}{T} \int^\infty_0 dr\ r^2 \left[ \frac{1}{2}\left( \frac{d\rho}{dr} \right)^2+ \tilde{\mathcal{F}}(\rho,T) \right],& T \gg R_0^{-1}
 \end{array}\right.
\end{equation}
In these limits, the equations of motion for the bounce become:
\begin{equation}  \label{eq:all_bounces}
 \frac{d^2\rho}{dr^2}+\frac{\alpha}{r}\frac{d\rho }{dr}-\frac{\partial \tilde{\mathcal{F}}}{\partial \rho}(\rho,T)=0, \qquad \left.\frac{d\rho}{dr}\right|_{r=0}=0, \qquad  \lim_{r\rightarrow \infty}  \rho(r)= v_{T}^{(+)}
\end{equation}
with $\alpha=2$ for $T \gg R_0^{-1}$ and $\alpha=3$ with $r$ replaced by $\tilde{r}$ for $T \ll R_0^{-1}$. The prefactor $A(T)$ in equation (\ref{eq:decay_prob_general}) also admit different forms in the low and high temperature limits:
\begin{equation}  \label{eq:all_prefactor}
A(T) \approx \left\{\begin{array}{ll}
  \displaystyle \frac{1}{R_0^4}\left(\frac{S_4(T)}{2 \pi}\right)^2&, T \ll R_0^{-1}  \\
 \displaystyle T^4\left(\frac{S_3(T)}{2 \pi T}\right)^{3/2}&, T \gg R_0^{-1}
  \end{array}\right.
\end{equation}
The difference in these expressions comes from the fact that the $O(4)$-symmetric bounce has 4 zero-modes contributing a factor $\left[S/(2 \pi)\right]^{1/2}$ each, while the $O(3)$-symmetric solution only has 3 zero-modes.

In the case of a rapid phase transition occurring around the electroweak scale $T_{EW}\sim 100$ GeV, the high-temperature formula provides a good approximation. However, it is not clear a priori how $T_p$ and $R_0$ will scale if the transition occurs with a significant amount of supercooling. In particular if $T_p \lesssim R_0^{-1}$, approximating $S$ by $S_3/T$ might not be accurate anymore, requiring the use of the exact expression (\ref{eq:S_action}) (or $S_4$ at even lower temperature). For this reason, we compare how each of the three different actions $S(T)$, $S_4(T)$ and $S_3(T)$ behaves as a function of the temperature. To do this, the bounce equations of motion must be solved numerically. In the low and high temperature regime, (\ref{eq:all_bounces}) is an ODE and can be integrated using the shooting method\footnote{Note that there is no bounce solution when the two vacua are exactly degenerate at $T_c$ and that tunnelling occurs only for $T<T_c$. Numerically, the shooting method provides solutions only for a wide enough energy separation between the vacua, namely for $T\leq T_{\star}<T_c$. Although tunnelling solutions can exist for $T_{\star}<T<T_c$ and be estimated through the thin-wall approximation, they are negligible for the PT as $\Gamma$ is more and more suppressed as the vacua are more and more degenerate (see e.g. Sec. IV in \cite{Coleman:1977py}). }. On the other hand, the spacetime dependent equation (\ref{eq:bounce_general}) is a PDE and thus more difficult to address. Following \cite{Ferrera:1995gs,Salvio:2016mvj}, we discretise the spacetime over a lattice. The PDE and the boundary conditions reduce then to a set of non-linear algebraic equations located at each point of the lattice. This set of equations is solved according to the Newton's method: starting from a guess solution we build a new solution which minimises the error and iterate until the error becomes small enough. For this method to converge, the choice of the guess is important. In our case, we use the zero-temperature $O(4)$ solution, found from the shooting method, as a guess to solve (\ref{eq:bounce_general}) at $T=0+\Delta T$. This solution is then used to solve the problem at $T^{(n+1)}= T^{(n)}+ \Delta T$ recursively. The numerical solutions will be presented in Sec.~\ref{sub:nume_solutions}.

\subsection{Bubble dynamics and energy}\label{sub:long_pt}
Given the nucleation probability $\Gamma(T)$ discussed in the previous section, we can now describe the dynamics of a first-order PT. We apply the general formalism provided in \cite{Turner:1992tz}. We consider an expanding Universe with scale factor $a(t)$ and Hubble rate $H=\dot{a}/a$. The probability $p(t)$ for a given point of spacetime to be in the symmetric phase at time $t$ is then given by \cite{Turner:1992tz}:
\begin{equation}\label{eq:prob_point_in_false_vacuum}
  \begin{aligned}
    p(t)= \exp \left[ -\underbrace{\frac{4\pi}{3} \int^t_{t_c} dt' \Gamma(t') a^3(t') r^3(t,t') }_{:=\mathcal{I}(t)}\right]
  \end{aligned},
\end{equation}
where $\mathcal{I}(t)$ corresponds to the volume occupied by the true vacuum bubbles\footnote{Note that the exponentiation of $\mathcal{I}$ in (\ref{eq:prob_point_in_false_vacuum}) corrects the fact that regions with overlapping bubbles have been counted twice in $\mathcal{I}$.}. Indeed, bubbles which have nucleated at $t'<t$ with probability $\Gamma(t')$ would have then grown until $t$ reaching a (coordinate) radius $r(t,t')$ given by:
\begin{equation}\label{eq:bubble_radius_general}
  \begin{aligned}
    r(t,t') = \int^t_{t'} dt''  \frac{v(t'') }{a(t'')},
  \end{aligned}
\end{equation}
with $v(t)$ being the bubble wall velocity. In the previous equation, we have neglected the initial radius of the bubble which rapidly becomes negligible compared to the expanding size.

The condition that the phase transition completes can be translated to the condition that $p(t)\to 0$ for $t>t_c$. As we are ultimately interested in the production of gravitational waves from bubble collisions, we are looking for the transition time corresponding to the period of maximum bubble collisions. This period can be estimated by the percolation time $t_p$ \cite{Leitao:2012tx,Leitao:2015fmj}. According to numerical simulations performed with spheres of equal size, percolation occurs when approximately $29\%$ of space is covered by bubbles \cite{Vinod}. As suggested by \cite{Leitao:2012tx,Leitao:2015fmj}, we thus define $t_p$ from the condition $p(t_p) \approx 0.7$. 

Knowing the collision time, we can then look for the distribution of number of bubbles at that time as a function of their size. From (\ref{eq:bubble_radius_general}), a bubble formed at time $t_R$ will have a physical size  $R(t,t_R)=a(t)r(t,t_R)$ at time $t$. The number density of such bubbles is then given by \cite{Turner:1992tz}:
\begin{equation}
\label{eq:bubble_number_density}
\frac{dN}{dR}(t,t_R)=\Gamma(t_R)\left(\frac{a(t_R)}{a(t)}\right)^4 \frac{p(t_R)}{v(t_R)}.
\end{equation}
For $t=t_p$, the peak of this distribution gives us the size $\bar{R}$ of the majority of the bubbles which are colliding. Equivalently, it also provides the time $t_{\bar{R}}$ when most of these bubbles have been produced. We call this moment the nucleation time $t_n$ (rather than $t_{\bar{R}}$) and it can be explicitly computed via:
\begin{equation}\label{eq:percolation_nucleation_temperatures}
\left.\frac{d}{dt_R}\left(\frac{dN}{dR}(t_p, t_R)\right)\right|_{t_R=t_n}=0.
\end{equation}
As we shall see in Sec.~\ref{sec:gw}, $\bar{R}:=R(t_p, t_n)$ is the key parameter to determine the peak frequency of the GW spectrum produced by bubble collisions.

Another important parameter, related to the amplitude of the GW spectrum, is the kinetic energy stored in the bubble walls. This kinetic energy comes from the vacuum energy released during the transition from the unbroken phase to the broken phase of the scalar field $\rho$. In order to derive this quantity, we briefly remind the basic thermodynamic properties of this field. As described above, its free energy is given by the effective potential, $\mathcal{F}(\rho, T)=V(\rho, T)$, and this allows us to define the pressure $p=-\mathcal{F}$ and the energy density $\epsilon(\rho, T)=\mathcal{F}-T \frac{d \mathcal{F}}{dT}$. The released vacuum energy density is associated with the following latent heat: $\tilde{\epsilon}(T)=\epsilon(v_{T}^{(+)}, T)- \epsilon(v_{T}^{(-)}, T)$. During the transition, this latent heat is converted into the formation of the bubbles (surface energy and kinetic energy of the walls) and into the reheating and fluid motion of the plasma. Following the notation of \cite{Caprini:2015zlo}, we write $\kappa_{\rho}$ the fraction of energy which goes into the kinetic energy of the bubbles (i.e. the scalar field $\rho$). In our case, we can assume that $\kappa_{\rho} \sim 1$ as we are considering a very strong phase transition (see the discussion below for more details). As a result, the kinetic energy of a bubble is given by $\tilde{\epsilon}$ and the portion of space it has converted. For bubbles produced at $t_n$, their kinetic energy at the percolation time $t_p$ is then
\begin{equation}
\label{eq:kinetic_energy_bubble}
E_{\text{kin}} =4 \pi \int_{t_n}^{t_p}dt \frac{dR}{dt}(t,t_n) R^2(t,t_n)\tilde{\epsilon}(t) 
\end{equation}
where we have taken into account the fact that the latent heat varies with time (namely temperature). In the case of a short phase transition or a slowly varying $\tilde{\epsilon}$, the above equation reduces to $E_{\text{kin}}=\frac{4 \pi}{3} \bar{R}^3 \tilde{\epsilon}$ as we should expect.

 %\begin{equation}
%\label{eq:latent_heat}
%\epsilon(T)= \left( V(v_{T}^{(+)}, T)-V(v_{T}^{(-)}, T)\right) + T \left( \frac{\partial V}{\partial T}(v_{T}^{(-)}, T) - \frac{\partial V}{\partial T}(v_{T}^{(+)}, T)\right).
%\end{equation}

In order to explicitly compute $\bar{R}$ and $E_{\text{kin}}$, we need to determine the bubble growth which depends on the velocity $v(t)$ and the scale factor $a(t)$ according to Eq. (\ref{eq:bubble_radius_general}). We discuss the details of the evolution of the background Universe in the next section. Regarding the velocity, it is usually a difficult task to calculate precisely $v(t)$ as its depends on the interaction between the bubble wall and the plasma. However, it has been shown that for phase transitions with a sufficient amount of supercooling the produced bubbles quickly reach the speed of light \cite{Caprini:2015zlo}. They are referred to as runaway bubbles. Indeed, the amount of converted vacuum energy is such that the energy deposited in the plasma saturates and the majority goes into accelerating the bubble wall. This also confirms the previous assumption that $\kappa_{\rho} \sim 1$. Note that this statement has been rigorously verified in \cite{Kobakhidze:2016mch} (see their Sec. 4) for our model of interest given in Sec.~\ref{sec:model}. 

\newpage

\subsection{Equation of state}
\label{sec:eos}

In order to carefully describe the dynamics of a long-lasting phase transition, the expansion of the Universe cannot be neglected and this requires to determine the scale factor $a(t)$. In the same way, it is also important to know how the temperature of the Universe, $T(t)$, evolves during the process. Both these quantities depend on the equation of state (EOS) of the different components of the Universe and which of them dominate. In particular, if the Universe is dominated by a single component with EOS $p=w \epsilon$, the scale factor is then given by $a(t)\propto t^\gamma$ with $\gamma =\frac{2}{3 (w+1)}$ ($w\neq -1$). For $w< -1/3$ ($\gamma >1$), it follows that the Universe undergoes an accelerated expansion (power-law inflation). In the same way, the case $w=-1$ (vacuum domination) also leads to an accelerating phase with $a(t)\propto e^{Ht}$ (exponential inflation).
 
In the general scenario of electroweak PT, bubbles nucleate near the electroweak scale, $T_{EW} \sim 100$ GeV, and percolate rapidly. During such a process, the Universe is radiation dominated with
\begin{equation}
\label{eq:radiation_dominated_universe}
p=\frac13 \epsilon, \qquad  a(t) \propto t^{1/2}, \qquad t= \underbrace{\left(\frac{45 M_{p}^2}{16 \pi^3 g_{\star}}\right)^{1/2}}_{:=\xi} \frac{1}{T^2},
\end{equation}
with $M_p\sim 1.22 \times 10^{19}$ GeV the Planck mass and $g_{\star}\sim 100$ the effective number of relativistic degrees of freedom in the symmetric phase. However, this equation of state might not be valid in the case of strong supercooling or for a prolonged transition. The reason comes from the fact that as the Universe cools down, the vacuum energy density of the scalar field which remains in the unbroken phase starts dominating over the radiation energy density, $\epsilon_{\text{rad}}= \pi^2 g_{\star} T^4/30$, possibly leading to the phase of inflation described above.

In order to have a general understanding of transitions which such a behaviour, we introduce the time $t_e$ of radiation-vacuum equality satisfying $\epsilon_{\text{vac}}(t_e)=\epsilon_{\text{rad}}(t_e)$.\footnote{For simplicity, we assume here that the Universe is dominated by a single component at a time and that the transition is sharp between radiation and vacuum domination.} In the standard case, $t_n\lesssim t_p \ll t_e$, and the vacuum energy is released into the bubbles before it could dominate. Then, scenarios with an inflationary background have been considered  for some classes of scale-invariant models in \cite{Randall:2006py,Konstandin:2010cd,Konstandin:2011dr}. In such cases, most of the bubbles nucleate after radiation-vacuum equality, namely $t_e <  t_n \lesssim t_p$. On the other hand, the process we want to describe in this article (prolonged PT) is different from the two previous ones in the sense that $ t_n <  t_e <t_p$, namely bubbles are produced before vacuum energy would dominate and percolation requires a long time to complete. We now address this type of transition in more details.

The large separation between nucleation and percolation comes from a decay probability $\Gamma$ weaker than in the standard case, such that less bubbles are produced per unit volume and more time is required for them to collide. Let us clarify this reasoning  by assuming that all bubbles are nucleated at $t_n$ such that $\Gamma(t)=\bar{\Gamma}(t_n)\delta(t-t_n)$. The exponent in Eq. (\ref{eq:prob_point_in_false_vacuum}) becomes $\mathcal{I}(t)=\frac{4 \pi}{3} \bar{\Gamma}(t_n) a^3(t_n) r^3(t, t_n)$ and this clearly shows how a larger radius (i.e. longer time) compensates for a weaker nucleation probability. However, this last expression also indicates than the transition might never complete if the Universe is in accelerating expansion, since in such a case $r(t,t_n)$ is bounded\footnote{This can easily be seen explicitly. Assuming $a(t)\propto t^{1/2}$ for $t<t_e$ and $a(t) \propto t^\gamma$ for $t>t_e$, Eq. (\ref{eq:bubble_radius_general}) (with $v\sim1$) gives $r(t,t_n)\propto 2(t_e^{1/2}-t_n^{1/2})+t_e^{\gamma-1/2}(t^{1-\gamma}-t_e^{1-\gamma})/(1-\gamma)$, such that when $t\to \infty$  $r(t,t_n)\to +\infty$ if $\gamma <1$ and $r(t,t_n)\to (2+1/(\gamma-1))t_e^{1/2}-2 t_n^{1/2}$ if $\gamma >1$. } when $t\to \infty$. In other words, there is the possibility for the bubbles to not grow fast enough in order to reach each other and to collide.

However, we now argue that such a scenario (with no percolation) is unlikely to occur as long as enough bubbles are produced during the radiation dominated period, namely before $t_e$. Indeed, as bubbles nucleate, vacuum energy is converted into kinetic energy of the wall motion such that the energy budget at the time $t_e$ is not simply dominated by vacuum energy density even if the bubbles have not yet collided. Actually, the bubbles are acting as inhomogeneity in the background of the expanding space-time and this render difficult to naively estimate what would be the corresponding dynamics of the Universe. According to several studies including numerical simulations \cite{Mazenko:1985pu,Goldwirth:1989pr,Goldwirth:1991rj} (see \cite{Brandenberger:2016uzh} for a recent review), it has been shown that small-field inflation is very unlikely to proceed with inhomogeneous initial conditions. We shall then assume in the following part of the article that for a sufficient number of bubbles produced at $t_n$, the Universe expansion will not accelerate around $t_e$ and that percolation does occur at a given time $t_p> t_e$.

An exact description of the evolution (namely a precise value of $\gamma$) would require numerical simulations which are beyond the scope of this study. As we expect no acceleration because of the previous argument, we have $\gamma<1$ and so we assume for simplicity that the Universe remains radiation dominated during the entire process ($\gamma = 1/2$). Deviation of the value of $\gamma$ in the range $[0,1]$ would change the estimation of the parameters describing the transition, in particular $\bar{R}$ and $E_{\text{kin}}$, but not the qualitative picture. Moreover, we expect such a deviation to be compensated by a shift in the value of the initial conditions describing the underlying particle physics model (the parameter $\kappa$ in Eq. (\ref{eq:potential_tree_level}) in our case).

Under the aforementioned assumptions ($\gamma \sim 1/2$, $v\sim 1$, $\kappa_{\rho}\sim 1$) and using Eq. (\ref{eq:radiation_dominated_universe}), we can simplify Eqs. (\ref{eq:prob_point_in_false_vacuum}), (\ref{eq:bubble_radius_general}) and (\ref{eq:kinetic_energy_bubble}) and write them in terms of temperature rather than time. Regarding the evolution of the temperature, we also recall that for a strong transition the dominant part of the vacuum energy is transformed into kinetic energy of the bubble walls meaning that we can neglect heating of the plasma. In the same way, the kinetic energy is subsequently transformed into GW energy through bubble collisions such that again heating is negligible. We eventually obtain the following key equations:
\begin{equation}\label{eq:prob_point_in_false_vacuum_temperature}
  \begin{aligned}  
   R(T, T') &= \frac{2 \xi}{T}\left(\frac1{T}-\frac1{T'}\right) \\
    p(T)&= \exp \left[ -\frac{64\pi}{3} \xi^4 \int_T^{T_c} dT' \frac{\Gamma(T')}{T'^6} \left(\frac1{T}-\frac1{T'}\right)^3 \right] \\
    E_{\text{kin}}&=32 \pi \xi^3 \int_{T_p}^{T_n} dT \frac1{T^3} \left(2-\frac{T}{T_n}\right)\left(\frac1{T^2}-\frac1{T T_n}\right)^2 \tilde{\epsilon}(T) ~.
  \end{aligned}
\end{equation}
It is now possible to numerically evaluate the previous expressions and to derive the key parameters $T_n$, $T_p$, $\bar{R}$ and $E_{\text{kin}}$ defining the phase transition. We present the results in the next section.

It is worth mentioning how the above formalism simplifies in the case of a quick phase transition, which is the main situation investigated in the literature. In that case, the PT is assumed to proceed rapidly around the temperature $\tilde{T}_n$ when at least one bubble has been produced per Hubble volume, namely $\int_{\tilde{T}_n}^{T_c}dT \frac{\Gamma(T)}{H^4(T) T} \sim 1$. In this context, $\tilde{T}_n$ is called the nucleation temperature and replaces our expression $T_n$ derived from Eq. (\ref{eq:percolation_nucleation_temperatures}). Then the decay probability can be expanded around that instant as $\Gamma(t) \approx \Gamma(\tilde{t}_{n}) e^{\beta(t-\tilde{t}_{n})}$, where $\beta^{-1}$ gives the time scale of the transition. As such a PT is not expected to proceed too far below the electroweak scale, we have $\Gamma(T) \approx A(T) e^{-S_3(T)/T}$ and hence:
\begin{equation}
\label{eq:beta_coeff}
\frac{\beta}{H(\tilde{T}_n)}=\tilde{T}_n \left.\frac{d}{dT} \left( \frac{S_3(T)}{T}\right)\right|_{T=\tilde{T}_n}.
\end{equation}
The characteristic size and energy of the bubbles are then expected to be $\tilde{R}=v\beta^{-1}$ and $\tilde{E}_{\text{kin}}=\frac{4 \pi}{3}\tilde{R}^3 \epsilon(\tilde{T}_n)$ respectively.

\subsection{Numerical solutions}\label{sub:nume_solutions}
We give the numerical results of the previous formalism applied to the model described by the potential (\ref{eq:potential_tree_level}) and (\ref{eq:finite_potential}). The range of the parameter $\kappa \in [-1.85,-1] \frac{m_h^2}{v}$ has already been investigated in \cite{Kobakhidze:2016mch}. In that case, the phase transition occurs quickly and can be described by a rapid phase transition (as explained in the last paragraph of the previous section). It results in the production of a GW spectrum potentially detectable by eLISA. However, for  $\kappa < -1.85 \frac{m_h^2}{v}$, the transition lasts longer and we require the more general prescription outlined in this paper.

\begin{figure}[h]
    \centering
    \includegraphics[scale=0.7]{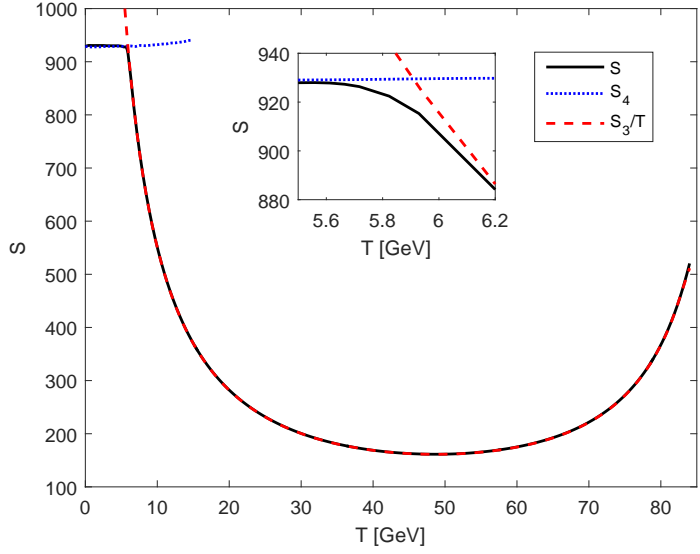}
     \caption{Thermal behaviour of the Euclidean action for $\kappa = -1.9\frac{m_h^2}{v}$. The black solid line corresponds to the action $S(T)$, Eq. (\ref{eq:S_action}), whose spacetime dependent bounce solution (\ref{eq:bounce_general}) has been solved over a lattice through Newton's method. The blue doted line (resp. red dashed line) is the low (resp high) temperature approximation $S_4(T)$ (resp. $S_3(T)/T$) given by Eq. (\ref{eq:all_actions}). }
     \label{fig:actions}
\end{figure}

\begin{figure}[h]
    \centering
    \includegraphics[scale=0.7]{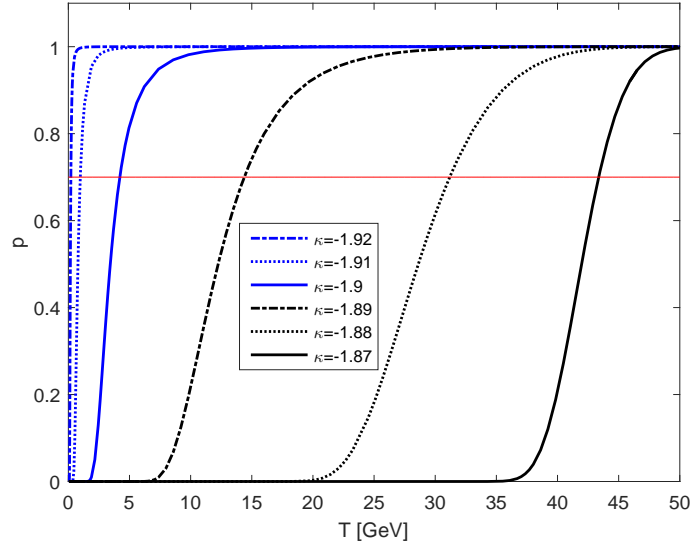}
     \caption{Thermal evolution of the probability $p(T)$ for $\kappa = [-1.92,-1.87]\frac{m_h^2}{v}$. The intersection between the curves and the red solid line $p(T)=0.7$ gives the percolation temperature.}
     \label{fig:probs}
\end{figure}

First, we show in Fig.~\ref{fig:actions} how the Euclidean action behaves as a function of temperature, for $\kappa = -1.9\frac{m_h^2}{v}$. It appears that the action $S(T)$ given by Eq. (\ref{eq:S_action}) is not only well approximated by $S_4(T)$ at $T\ll R_0^{-1} \approx 6$ GeV and $S_3(T)/T$ at $T\gg R_0^{-1}$, but is also close to $R_0^{-1}$. In other words, the $\min\{S_4,S_3/T\}$ provides a good approximation of the action over the entire range of temperatures considered, as also suggested in \cite{Linde:1981zj}. This observation is important from the computational point of view as it means we can avoid solving the time-dependent PDE (\ref{eq:bounce_general}) which is computationally expensive. Moreover, we observe that the action becomes, and stays, large at low temperatures ($S\sim S_4 \sim 930$), meaning that its effect on the PT dynamics is exponentially suppressed (see for example (\ref{eq:decay_prob_general})).

Second, $p(T)$ is computed from $\Gamma(T)$ according to Eq. (\ref{eq:prob_point_in_false_vacuum_temperature}). The results for several key values of $\kappa$ are given in Fig.~\ref{fig:probs}. As expected, the phase transition can be identified as a rapid change in $p(T)$ from $1$ to $0$. The corresponding nucleation and percolation temperatures are given in Table \ref{tab:alpha_beta}. We observe that $T_n\sim 49$ GeV for each $\kappa$. This is due to the fact that most of the bubbles are produced when the action $S_3(T)/T$ reaches its minimum, whose location only slightly changes with $\kappa$. On the other hand, we notice that $T_p$ varies through several order of magnitudes. This is because the number density of bubbles produced at the nucleation time changes as a function of $\kappa$. This confirm the expectation that as the decay probability decreases, more time is needed for the transition to complete. We have verified that these results are consistent with our assumption that most bubbles are produced before vacuum energy dominates. Indeed, vacuum-radiation equality would occur at $T_e \sim 35.5$ GeV for this range of $\kappa$, if no bubbles were produced earlier. This confirms $T_n > T_e$.

The remaining task is to compute the characteristic bubble size $\bar{R}$ and the kinetic energy $E_{\text{kin}}$ of the bubbles at the percolation temperature. For later convenience, we rescale them compared to the Hubble radius and radiation energy density at this time. To this end, Table \ref{tab:alpha_beta} uses the dimensionless parameters $(\bar{R}H_p)^{-1}$ and $\alpha=\epsilon_{\text{kin}}/ \epsilon_{\text{rad}}(T_p)$ with $\epsilon_{\text{kin}}=E_{\text{kin}}/\bar{R}^3$. In this way, one can easily compare our results with those mentioned in the literature (as $v\sim 1$, $(\bar{R}H_p)^{-1}$ takes the role of $\beta/H$ given by Eq. (\ref{eq:beta_coeff})).\footnote{Note that Eq. (\ref{eq:beta_coeff}) could clearly not have been applied in this scenario as $\beta$ would have been negative for temperature below the minimum of $S_3/T$.} 
 
\begin{table}[t]
\begin{center}\renewcommand{\arraystretch}{1.3}
\begin{tabular}[c]{c c c c c c }
\hline
$\kappa$ $[m_h^2/|v|]$& $T_c$ GeV & $T_n$ GeV & $T_p$ GeV & $(\bar{R}H_p)^{-1}$ & $\epsilon_{\text{kin}}/ \epsilon_{\text{rad}}$ \\ \hline
\rowcolor{gray!20}
$-1.87$ & 98.3 & 48.9 & 43.4 &  8.79 & 0.57 \\ \hline 
$-1.88$ & 98.0 & 48.9 & 31.2 &  2.76 & 1.88 \\ \hline 
\rowcolor{gray!20}
$-1.89$ & 97.7 & 49.0 & 14.4  & 1.41 & 37.8 \\ \hline 
$-1.9$ & 97.4 & 48.7 & 4.21  &  1.09 & $5.09\cdot10^3$ \\ \hline 
\rowcolor{gray!20}
$-1.91$ & 97.1 & 48.6 & 0.977  & 1.02 &  $1.73\cdot10^6$ \\ \hline 
$-1.92$ & 96.8 & 48.5 & 0.205  & 1.00 &  $8.80\cdot10^8$ \\ \hline 
\end{tabular}
\caption{Key parameters describing the phase transition for $\kappa = [-1.92,-1.87]\frac{m_h^2}{v}$. This range of $\kappa$ values has been selected in prevision of its relevance for GW production. \label{tab:alpha_beta}}
\end{center}
\end{table}

We confirm from Table \ref{tab:alpha_beta}, that there exists some parameters for which the PT completes well below the electroweak scale. Lower values of percolation temperature are also possible for lower values of $\kappa$, but they are constrained from various considerations. First, the obvious lower bound $T_p \gtrsim 1$ MeV is given by nucleosynthesis constraints. Second, we shall see in the next section that the GW spectrum for $\kappa \lesssim -1.92 \frac{m_h^2}{v}$ is already excluded from current PTA surveys. \newline
Note also that at temperatures around $T \sim 100$ MeV, the QCD phase transition, which is believed to be second-order, takes place. As discussed in \cite{Witten:1980ez}, a $q\bar{q}$ condensate with non zero vacuum expectation value is thus expected to form and to contribute to the thermal potential (\ref{eq:finite_potential}) via a linear term in the Higgs field. This effect has no significant influence on our model since the cubic term in the tree-level potential (\ref{eq:potential_tree_level}) induces a large barrier which is not affected by such a linear term. It is however of importance for models in which the barrier becomes weaker at low temperature, see e.g. \cite{Witten:1980ez,Buchmuller:1990ds,Arunasalam1}. \newline
Another observation from Table \ref{tab:alpha_beta}  is that the two quantities $(\bar{R}H_p)^{-1}$ and $\frac{\alpha}{\alpha+1}$ (which will be important in the next section) approach $1$ for lower and lower values of $\kappa$. The fact that $(\bar{R}H_p)^{-1} \to 1$ means that bubbles are almost of horizon size when they collide and that they never become of super-horizon size because $H^{-1}$ also increases linearly with time. Note also that the increase in $\alpha$ for lower $\kappa$ is mainly due to $\alpha \propto T_p^{-4}$ rather than to the change of kinetic energy stored in the bubbles.

%%%%%%%%%%%%%%%%%%%%%%%%%%%%%%%%%%%%%%%%%%%%%%                       
%%Section
%%%%%%%%%%%%%%%%%%%%%%%%%%%%%%%%%%%%%%%%%%%%%%
\section{Gravitational wave production}\label{sec:gw}

A stochastic background of gravitational waves is usually described in terms of its contribution to the energy density of the Universe per frequency interval:
\begin{equation}
  \label{eq:gw_density_over_frequency}
  h^2 \Omega_{GW}(f) = \frac{h^2}{\epsilon_c} \frac{d \epsilon_{GW}}{d (\ln f)},
\end{equation}
where $f$ is the frequency, $\epsilon_{GW}$ the gravitational wave energy density,  and $\epsilon_c= 3H_0^2/(8\pi G) $ is the critical energy density today. The production of GWs from a first-order phase transition originates from three sources: the collisions of bubbles walls \cite{Turner:1990rc,Kosowsky:1991ua,Kosowsky:1992rz,Kosowsky:1992vn,Kamionkowski:1993fg,Huber:2008hg,Caprini:2007xq}, sound waves in the plasma formed after collision \cite{Hindmarsh:2013xza,Hindmarsh:2015qta,Giblin:2013kea,Giblin:2014qia} and magnetohydrodynamics turbulences in the plasma \cite{Caprini:2006jb,Kahniashvili:2008pf,Kahniashvili:2008pe,Kahniashvili:2009mf,Caprini:2009yp}. As these three contributions should approximately linearly combine \cite{Caprini:2015zlo}, the total energy density can be written as
\begin{equation}\label{eq:gw_density_decomposition}
  h^2 \Omega_{\text{GW}} \simeq h^2 \Omega_{col} + h^2 \Omega_{sw} +h^2 \Omega_{\text{MHD}}~. 
\end{equation}
For models with a significant amount of supercooling (with $v\sim 1 $ and $\kappa_{\rho} \sim 1$ as described previously), the collision term $\Omega_{col}$ is dominant \cite{Caprini:2015zlo}. In this article, we can then assume that: 
\begin{equation}\label{eq:gw_density_higgs_field}
  h^2 \Omega_{\text{GW}} \simeq h^2 \Omega_{col}.
\end{equation}
We can thus expect the GW spectrum to be mainly produced around the percolation temperature $t_p$ of the phase transition with a characteristic frequency $f_p$. The amplitude of this signal then decreases as $a^{-4}(t)$ up to today while its frequency redshifts as $a^{-1}(t)$. In other words, the energy density stored in the GWs and the peak frequency today are given by \cite{Kamionkowski:1993fg}:
\begin{equation}\label{eq:redshift}
\arraycolsep=1.4pt\def\arraystretch{2.0}
\begin{array}{cll}
f_0 &= \displaystyle f_p \frac{a(t_p)}{a(t_0)} 
    &=  \displaystyle 1.65\times 10^{-7} \ \text{Hz} \left(\frac{f_p}{H_p}\right)  \left( \frac{T_p}{1\ \text{GeV}} \right)
           \left( \frac{g_*}{100} \right)^{1/6} \\
  \Omega_{GW,0} &= \displaystyle\Omega_{col} \left( \frac{a(t_p)}{a(t_0)} \right)^4 \left( \frac{H_p}{H_0} \right)^2    & \displaystyle= 1.67\times 10^{-5} h^{-2}\left( \frac{100}{g_*} \right)^{1/3} \Omega_{col}.
 \end{array}
\end{equation}

We can estimate the properties of the above GW spectrum from dimensional analysis. We expect the peak frequency to scale with the inverse size of the bubbles at their collision, namely $ f_p \sim (\bar{R})^{-1}$. Regarding the amplitude, we first have $\Omega_{col}=\epsilon_{GW}/\epsilon_{tot}$ where $\epsilon_{tot}$ is the total energy density of the Universe at the percolation time. The energy density of gravitational waves is then given by \cite{Maggiore:1999vm}: $\epsilon_{GW}=\frac1{8 \pi G} \langle \partial_t h_{\mu \nu} \partial_t h^{\mu \nu} \rangle$, where $h_{\mu \nu}$ is the metric perturbation. It satisfies Einstein's equations $\partial_{\alpha}\partial^{\alpha} h_{\mu \nu} \sim 8 \pi G T_{\mu \nu}$ with $T_{\mu \nu}$ the energy momentum tensor describing the source. In our case, we expect that $\partial_{\alpha} \sim \bar{R}^{-1}$ (the characteristic size of bubbles) and $T_{\mu \nu} \sim \epsilon_{\text{kin}}$ (the kinetic energy density stored in the bubbles). This implies $\epsilon_{GW} \sim 8 \pi G \bar{R}^2 \epsilon^2_{\text{kin}}$. Substituting $G$ from the Friedmann equation $H_p^2 = \frac{8 \pi G}{3} \epsilon_{tot}$, we get:
\begin{equation}
\label{eq:gw_spectrum_dimensional}
 \Omega_{col} \sim  ( \bar{R} H_p)^2 \frac{\epsilon_{\text{kin}}^2}{\epsilon_{tot}^2} \sim  ( \bar{R} H_p)^2 \frac{\alpha^2}{(1+\alpha)^2}.
\end{equation}
where in the last equality we have used the fact that $\epsilon_{tot}=\epsilon_{\text{kin}}+\epsilon_{\text{rad}}$. As expected, the GW amplitude can then be estimated from the two parameters $\bar{R}H_p$ and $\alpha$ which have been computed in the previous section and are given in Table \ref{tab:alpha_beta}. Note that in general Eq. (\ref{eq:gw_spectrum_dimensional}) would also depend on $v$ and $\kappa_{\rho}$, which again have been assumed to be close to unity in our case.

Several studies provide more accurate expressions for the GW spectrum from bubble collisions, beyond the simple dimensional analysis. They usually rely on the envelope approximation and numerical simulations \cite{Kosowsky:1992vn,Kamionkowski:1993fg,Huber:2008hg}, although some analytical formula have also been suggested \cite{Caprini:2007xq,Jinno:2016vai}. Using the results of \cite{Huber:2008hg}, the spectrum today can be described as follows:
\begin{equation}\label{eq:coll}
  h^2 \Omega_{col}(f) =  1.67\times 10^{-5} 
  \left( \frac{100}{g_*} \right)^{1/3} \left(\frac{\beta}{H_p} \right)^{-2} \kappa_{\rho}^2 \left( \frac{\alpha }{1+\alpha} \right)^2 \left( \frac{0.11v^3}{0.42+v^2} \right)
S(f),
\end{equation}
where: 
\begin{equation}\label{eq:coll2}
  \begin{aligned}
    S(f)& =\frac{3.8 (f/f_0)^{2.8}}{1+2.8(f/f_0)^{3.8}},\\ 
    f_{0} &= 1.65 \times 10^{-7} \left( \frac{T_p }{1 \text{ GeV}} \right)\left( \frac{g_*}{100} \right)^{1/6}  \left( \frac{\beta}{H_p} \right) \left( \frac{0.62}{1.8-0.1v +v^2} \right)\ \text{Hz}.
  \end{aligned}
\end{equation}
It is important to realise that these formula have been derived under the assumption of a short-lasting PT as described at the end of Sec.~\ref{sub:long_pt}. This is why they use the coefficient $\beta$, defined by Eq. (\ref{eq:beta_coeff}), which corresponds to the time scale of the transition. In this article, we shall instead substitute $\beta\sim v \bar{R}^{-1} \sim \bar{R}^{-1}$ in Eq. (\ref{eq:coll}-\ref{eq:coll2}) in agreement with the dimensional estimate given by Eq. (\ref{eq:gw_spectrum_dimensional}). However, it is not clear if these fitted formula accurately describe a long-lasting transition. Obtaining a more precise result would require to further numerical simulations of bubble collisions without the assumption that the PT completes in a time less than the Hubble time. Such a task is beyond the scope of this article. Indeed, our main aim is to find in which broad range of frequency and amplitude lies the stochastic GW background produced by a supercooled and long-lasting transition and if this significantly deviates from the usual case. If so, a more precise description of the GW spectrum would be interesting to compute in future works. 

\begin{figure}[t]
    \centering
    \includegraphics[scale=0.5]{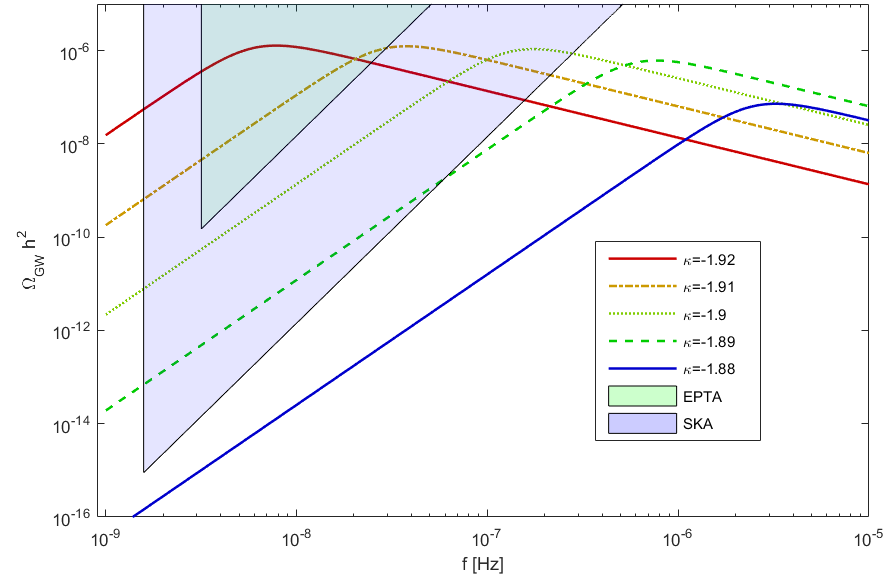}
     \caption{Lines: gravitational wave spectra estimated from Eqs. (\ref{eq:coll}-\ref{eq:coll2}) and the parameters of Table \ref{tab:alpha_beta} for $\kappa = [-1.92,-1.88]\frac{m_h^2}{v}$. Green shaded area: current exclusion limits from EPTA \cite{vanHaasteren:2011ni}. Blue shaded area: planned detection sensitivity of SKA  \cite{Moore:2014lga}.}
     \label{fig:gw_spectrum}
\end{figure}

It is now straightforward to evaluate Eqs.~(\ref{eq:coll})-(\ref{eq:coll2}) for the parameters given in Table~\ref{tab:alpha_beta}, describing the phase transition occurring in our model of interest. It appears that the peak frequency $f_0$ can be as low as $\sim 10^{-9}-10^{-7}$ Hz and thus lies in the detection range of PTA experiments. In Fig.~\ref{fig:gw_spectrum}, we compare the GW spectrum for several values of $\kappa$ and the current status of PTA detectors. Three collaborations have published limits on the amplitude of a stochastic GW background: the European Pulsar Timing Array (EPTA) \cite{vanHaasteren:2011ni}, the Parkes Pulsar Timing Array (PPTA) \cite{Shannon:2013wma} and the North American Nanohertz Observatory for Gravitational waves (NANOGrav) \cite{0004-637X-762-2-94}. All these three limits are of similar amplitudes and thus we only display the EPTA results\footnote{Note that the exclusion line is computed as explained in \cite{Moore:2014lga}.}. The sensitivity area should be improved in the future by the Square Kilometre Array (SKA)  \cite{5136190} whose expected detection range is also given in Fig.~\ref{fig:gw_spectrum} \cite{Moore:2014lga}. 

It is clear that our investigated model predicts GWs detectable by the aforementioned detectors and consequently we argue that this demonstrates a new method for probing first-order electroweak PTs. We might be tempted now to define the range of $\kappa$ values which is not yet excluded by EPTA and which could be tested by SKA. Although Fig.~\ref{fig:gw_spectrum} gives us a rough estimate around $\kappa \sim -1.9\frac{m_h^2}{v}$, this task would require a more precise computation of the GW spectrum taking into account the long duration of the transition in the numerical simulations. We also remind that these results rely on our assumption of a radiation dominated period. As explained earlier, the exact equation of state describing the Universe during the transition is expected to be more complicated with the potential effect of changing the value of the parameters entering in the GW spectrum (but not the general behaviour). However, we emphasize that the model we discussed in Sec.~\ref{sec:model} was mainly introduced as a case of study to illustrate the dynamics of a supercooled and long-lasting PT. We expect other theories to present similar features, especially a class of scale-invariant models with very light scalar particles \cite{Witten:1980ez,Kobakhidze:2017eml,Arunasalam1}.

%%%%%%%%%%%%%%%%%%%%%%%%%%%%%%%%%%%%%%%%%%%%%%
%%Section
%%%%%%%%%%%%%%%%%%%%%%%%%%%%%%%%%%%%%%%%%%%%%%
\section{Conclusion}\label{sec:conclusion}
In this article, we investigated the production of gravitational waves during a strongly supercooled electroweak phase transition. Considering a particle physics model based on a nonlinear realisation of the electroweak gauge group, we carefully computed the dynamics of the Higgs field during such a transition. In particular, we found an interesting range of parameters for which the PT completes at a temperature significantly below the electroweak scale. This analysis required to take into account the expansion of the Universe and to study the behaviour of the nucleation probability of true vacuum bubbles at low temperature. Regarding the latter point, we compared the usual low and high temperature expressions of the Euclidean action to a more general formula based on a time-dependent bounce solution. We observed that the two approximate equations actually provide a good estimate of the action over the entire range of valid temperatures.

In this scenario, we argued that GWs are produced by the collisions of large bubbles (of the order the horizon size) with a kinetic energy density significantly higher than the radiation energy density of the Universe at the time of collision. This results in a large amplitude stochastic GW background in the frequency range $10^{-9}-10^{-7}$ Hz which can be probed by pulsar timing arrays. We derived this prediction from both dimensional arguments and the use of previous numerical simulations of colliding bubbles. Although it is clear from these analysis that our model of interest predicts a GW spectrum in the sensitivity band of PTA detectors, more refined simulations would be needed to improve the accuracy of our results. In particular, a better estimation of the exact equation of state and scale factor during the phase transition would increase the accuracy of the relation between GW predictions and specific values of the parameter $\kappa$ of the particle physics model we considered.

In summary, we showed that a sufficiently supercooled electroweak phase transition can be detected with pulsar timing arrays. This enlarges the way of probing first-order cosmological PT in addition to previous proposals with space-base interferometers such as eLISA.  This also increases the prospects of new generation PTA detectors like SKA. 

%%%%%%%%%%%%%%%%%%%%%%%%%%%%%%%%%%%%%%%%%%%%%%
%%Acknowledgements
%%%%%%%%%%%%%%%%%%%%%%%%%%%%%%%%%%%%%%%%%%%%%%
\acknowledgments
We would like to thank Csaba Balazs and Alexander Kusenko for useful discussions. This work was partially supported by the Australian Research Council.

%%%%%%%%%%%%%%%%%%%%%%%%%%%%%%%%%%%%%%%%%%%%%%
%%Appendix
%%%%%%%%%%%%%%%%%%%%%%%%%%%%%%%%%%%%%%%%%%%%%%
\clearpage
\appendix
\section{Appendix}
%%%%%%%%%%%%%%%%%%%%%%%%%%%%%%%%%%%%%%%%%%%%%%
%Section
%%%%%%%%%%%%%%%%%%%%%%%%%%%%%%%%%%%%%%%%%%%%%%
\subsection{Finite Temperature Potential}\label{app:fin_temp_pot}
We give here the detailed expression of the finite temperature potential (\ref{eq:finite_potential}). The Coleman-Weinberg contribution $V_{CW}^{(1)}$ at $T=0$ is given by:
\begin{equation}
        V_{CW}^{(1)} (\rho) = \sum_{i=W,Z,t,h} n_i \frac{m_i^4(\rho)}{64 \pi^2}\left(\ln\left(  \frac{ m^2_i(\rho)}  {v^2}\right)- \frac{3}{2} \right) ~.
        \end{equation}
The finite temperature part $V^{(1)}(\rho,T)$ is defined via the thermal function $J$:
\begin{equation}\label{eq:thermal}
  \begin{aligned}
    V^{(1)}(\rho,T)&= \frac{T^4}{2\pi^2} \sum_{i=W,Z,t,h} n_i J\left[\frac{m_i^2(\rho)}{T^2}\right],\\
    J[m^2_i\beta^2] & :=\int^\infty_0 dx \ x^2 \ln \left[1-(-1)^{2s_i+1}e^{-\sqrt{x^2+\beta^2m^2_i}}\right],
      \end{aligned}
\end{equation}
where $s_i$ corresponds to the spin and $n_i$ to the number of degrees of freedom of the particle species $i$. The field-dependent masses $m_i(\rho)$ are given by
\begin{equation}
\begin{aligned} 
\label{eq:dof_mass_1}
 n_h&= 1,  \hspace{20ex}    &m_h^2(\rho)   =&  3{\lambda}\rho^2+ 2\kappa\rho - \mu^2 \\
   %\quad  m_\chi^2 (\rho) ={\lambda}\rho^2 - \mu^2 \\
 n_Z &=  3,   &m_Z^2 (\rho)  =&  \frac{g^2_2+g^2_1}{4}\rho^2, \\
 n_W&=  6, &m_W^2 (\rho)   =&  \frac{g^2_2}{4}\rho^2, \\
 n_t&= -12,   &m_t^2 (\rho)  = &\frac{y_t^2}{2}\rho^2.  
\end{aligned}
\end{equation}

The correction from Daisy terms can be described by a shift in the longitudinal components of the respective boson masses by their Debye correction (cf. e.g. \cite{Arnold:1992rz,Espinosa:1993bs}):
\begin{equation}
  \begin{aligned}
    m_h^2 &\rightarrow m_h^2 (\rho,T)&&=m_h^2(\rho)+\frac{1}{4} \lambda T^2+ \frac{1}{8}g_2^2 T^2 +  \frac{1}{16}(g_2^2+g^2_1) T^2+\frac{1}{4}y_t^2T^2,\\
    m_{W_L}^2 (\rho)&\rightarrow m_{W_L}^2(\rho,T) &&=m_{W}^2 (\rho)+\frac{11}{6}g_2^2T^2,\\
    m_{Z_L}^2 (\rho)&\rightarrow m_{Z_L}^2(\rho,T) &&= \frac{1}{2}\left[ m_Z^2(\rho) + \frac{11}{6}\left(g_2^2+g_1^2\right)T^2 + \Delta(\rho,T) \right],\\
    m_{\gamma_L}^2 (\rho)&\rightarrow m_{\gamma_L}^2(\rho,T)& &= \frac{1}{2}\left[ m_Z^2(\rho) + \frac{11}{6}\left(g_2^2+g_1^2\right)T^2 - \Delta(\rho,T) \right],\\
      \end{aligned}
\end{equation}
where:%{\color{red}[need verifiy from 1404.7673,1504.05949,9301285]}
\begin{equation}
  \Delta^2(\rho,T):=%m_Z^4(\rho) + \frac{11}{3}\frac{\cos^2 2\theta_W}{\cos^2 \theta_W}T^2\left[ m_Z^2(\rho)+ \frac{11}{12}\frac{g^2}{\cos^2 \theta_W}T^2 \right]
  \left( m_Z^2(\rho)+ \frac{11}{6}(g_2^2+g_1^2)T^2 \right)^2 -g_1^2g_2^2\frac{11}{3}T^2\left(\frac{11}{3}T^2+\rho^2  \right). 
\end{equation}
The number of degrees of freedom is then:
\begin{equation}
  g_{W_L}= 2g_{Z_L} = 2g_{\gamma_L}=2, \qquad g_{W_T}= 2g_{Z_T} =2 g_{\gamma_T}=4. 
\end{equation}

\bibliographystyle{mybibsty}
\bibliography{myrefs}

\end{document}